# Effective memory of the minority game

Chia-Hsiang Hung and Sy-Sang Liaw

*Department of Physics, National Chung-Hsing University, 250 Guo-Kuang Road, Taichung, Taiwan*

## Abstract

It is known that the memory is relevant in the symmetric phase of the minority game. In our previous work we have successfully explained the quasi-periodic behavior of the game in the symmetric phase with the help of the probability theory. Based on this explanation, we are able to determine how the memory affects the variance of the system in this paper. By using some particular types of fake history such as periodic type and random type, we determine how efficient the memory has been used in the standard game. Furthermore, the analysis on the effective memory strongly supports the result we proposed previously that there are *three* distinct phases in the minority game.

## I. Introduction

The minority game (MG)[1] is a simple model that can capture the essence of some features of the financial markets. The game is played by *N* agents in choosing one of two decisions. Those who are in the minority are rewarded one point. Suppose these *N* agents play the game many times, what outcome would be expected? Is there any statistical pattern? Can one of the agents manage to gain most? To answer these questions we need to take data from a MG played by real people which waits to be done[2]. Current study of the MG is to simulate the game on computers by assuming that each agent has a given type of strategy in choosing a decision. The standard MG [1] assumes that at each time, the winning decision of the last *M* time steps is known to each agent and each agent uses this information to make his decision according to one of his *S* strategies set up in the beginning of the game. A *strategy* is nothing but a list of all possible pieces of information with an arbitrary decision assigned to each of them. The parameter *M* is called the *memory*. A common piece of information consists of the winning decisions of last *M* time steps is called a *history*. There are $P = 2^M$ possible histories in a strategy. Agents evaluate their strategies by giving a virtual point to all strategies that have matched the winning decision at a time step, and use the best-scored one among his *S* strategies to make decision.

As one simulates the standard MG many times, the average of the number of winning agents is found to be, as expected, *N*/2. The variance per agent of winning population $\chi$, on the other hand, has an interesting dependence on the parameter $\alpha = P/N$ [3] that attracts many researchers. Cavagna[4] claimed that the global properties of $\chi$ is irrelevant to the information of the winning history by simulating the game with a random history at each time step and found the results were the same. Further quantitative studies[5] showed that the variance were not quite the same when different histories (winning history or random history) were used. Recently, Ho et al[6] has carried out a large scale numerical simulation to calculate $\chi$ in the $\alpha \ll 1$ region, called symmetric phase[7,8]. Results of using winning and random histories differ substantially in the symmetric phase.

People have found the behavior of the MG very interesting in the symmetric phase. In particular, the winning population changes quasi-periodically with a period of 2*P*[9,10,11]. It is now understood[12] that when $\alpha$ is very small, all strategies will be almost guaranteed to gain the same (*P*) points at the end of time step 2*P* so that at the time step $2P+1$ no strategy is preferred and the system is like going back to time step 1. Liaw et al[13] has calculated analytically the critical value $\alpha_{c_1}$ such that in the region $\alpha < \alpha_{c_1}$ the quasi-periodic behavior would appear. They employed a simple series of histories to explain the quasi-periodic behavior and calculated $\alpha_{c_1}$. We will introduce this series of histories in Sec. II and compare its values of $\chi$ with results of using true histories. We define effective memory in Sec. III and then calculate it for different types of rules in the symmetric phase. Sec. V is conclusion.

## II. Three kinds of history-updating rules

Let two possible decisions be 0 and 1. A history is then an *M*-digit binary number. At each time step, in the standard game the history is updated by dropping its first digit and attaching the winning decision to the end. We call this the standard updating rule for the history. In the standard updating rule, exactly two histories can be updated to each history, and each history can be updated to exactly two possible histories. A de Bruijn graph shows all the paths of the standard updating rule(Fig. 1a). One can see from Fig. 1a there are many loops in a de Bruijn graph. If instead an arbitrary history is given at each time step, we call it the random updating rule. Any history can follow a history with probability $1/P$. Random rule has been used by Caragna[4] to show the variation of $\chi$ with respect to $\alpha$ is qualitatively the same with that of using standard rule. Challet et al[14] has developed an exact solution for MG using random rule. A third updating rule, introduced by Liaw et al[12], is all possible histories take turns to appear regardless of what winning decisions are. If we order all possible histories by their decimal values, the rule can be arranged simply as history 0 is followed by history 1, and history 1 followed by 2, etc., and history *P* followed by 0. We call this the periodic updating rule whose path has only one loop(Fig. 1b). The periodic rule is simple and suitable for analytic analysis[12,13]. It can also be extended to study so called Thermal minority game[15]. Other updating rules are also available, but we will consider only the standard, random, and periodic rules in this paper.

Fig. 2 is a plot of typical results of variance per agent $\chi$ as a function of $\alpha$ using the standard, random, and periodic rules respectively. We see that results of these three rules are qualitatively alike. We separate the plot into three regions by two dotted line: $\alpha = \alpha_{c_1}$ and $\alpha = \alpha_{c_2}$. $\alpha_{c_2}$ is defined as the critical value of $\alpha$ at which $\chi$ has its minimum. It is not clear whether $\alpha_{c_2}$ has a common value for three different rules. For $\alpha$ larger than $\alpha_{c_2}$, $\chi$ monotonically increases and approaches ¼ which is the result of a random system—every agent uses random strategy at each time. For $\alpha$ smaller than $\alpha_{c_1}$, $\chi$ is larger than ¼ and approximately proportional to $1/\alpha$ [13]. In this region the values of $\chi$ depend on the rules strongly. In the second region, as $\alpha$ increases from $\alpha_{c_1}$ to $\alpha_{c_2}$, the value of $\chi$ decreases, drops below ¼ and reaches its minimum. Behavior of the system in this transition region is very interesting but hardly studied. In what follows we will focus on the symmetric phase $\alpha < \alpha_{c_1}$ only.

### III. Definition of effective memory

It is known that the system shows a quasi-periodic behavior in the first region $\alpha < \alpha_{c_1}$ no matter what rule is used. From Fig. 2 we see that the log-plot of $\chi$ versus $\alpha$ are roughly parallel in the first region and the order of the values of $\chi$ is $\chi_{rd} < \chi_{st} < \chi_{pd}$ for a given $P$. We may consider the $\chi$ versus $\alpha$ curve as a demand curve in microeconomics. When the demand decreases, the curve shifts to the left. Thus if we define the demand (of information) as the *effective information* $P'$, or equicalently, *effective memory* $M' = \log_2(P')$ of the game, we expect $P'_{rd} < P'_{st} < P'_{pd}$ and $\chi$ (the price) will be a monotonic increasing function of $P'$.

To find a suitable definition for $P'$, we simply construct a relation between $\chi$ and $P'$. As we play the game again and again, when a history appears a second time, the path of the histories—the history and all histories between its two consecutive appearances, forms a loop. It was found[16,12] for small $\alpha$ case that when a loop is found, or equivalently, when a history appears a second time, the winning decision is most likely be different from that of the last time when the history appeared. The quasi-periodic behavior is result of this property. In Ref. [12] we have shown that when a history appears a second time, the population variance will shift away from 0 by a value which is inversely proportional to the number $l$ of histories that have appeared odd times. And the average variance per agent over a quasi-period $2P$ is given by[13]

$$\chi = \frac{1}{8} + \frac{c_S}{\alpha} \sum_l \frac{1}{l} \tag{1}$$

where $c_S$ is a constant dependent on the number of strategies $S$, and 1/8 comes from the contributions of $P$ time steps when system behaves in random fashion. The summation is over all loops. Let us consider the case of using periodic rule first. At time step $P+1$, the history that appears in the time step 1 appears the second time, and all $P$ histories have appeared once right before the loop is closed. At time step $P+2$, the second history appears the second time, and $P-1$ histories have appeared once. In the same manner, the number $l$ of histories that have appeared once when a loop is just to be closed decreases by 1 in the subsequent time steps. At time step $2P$, $l$ is equal to 1. Consequently, average variance per agent over a quasi-period $2P$ is given by

$$\chi_{pd} = \frac{1}{8} + \frac{c_S}{\alpha} \sum_{l=1}^{P} \frac{1}{l} \approx \frac{1}{8} + \frac{c_S}{\alpha} (\log P + \gamma) \tag{2}$$

where $\gamma$ is Euler's constant: $\gamma = 0.5772$. For fixed $P$, $\chi_{pd} - 1/8$ is proportional to $\alpha^{-1}$, as expected from numerical simulation. For fixed $\alpha$, $\chi_{pd} - 1/8$ is proportional to $\log P + \gamma$. Because the periodic rule is the simplest rule for analysis, we use it as a base for comparison. That is, we define $P'_{pd} = P$. We then calculate $\chi_{rd}$ and $\chi_{st}$ and assume they have the form Eq. (2) with the effective information $P'_{rd}$ and $P'_{st}$ respectively.

### IV. Calculation of the effective memory

As we play the game again and again, there will be loops in the path of histories. Let us define the time-step difference between two consecutive appearances of a history the *length* of the loop. A longer loop means there are more histories have appeared odd times when the end of the loop is reached. And we have seen in the last section that $\chi$ is affected by the number of histories have appeared odd times when a particular history appears a second time. Therefore, the knowledge of the distribution of the lengths in a game is important to understand the variance.

When we use the periodic rule to update the history in playing the game, it is very easy to see that all loops have the same length, which is simply equal to $P$. If either the standard rule or random rule is used, the lengths of loops vary. We plot the frequency versus length ranging from 1 to $2P$ in Fig. 3 by playing the game $2P$ time steps and taking average of 650 runs. The length-distribution plot for the case of random rule can be calculated as follows. The probability of length 1 is $1/P$ because there is only one choice out of $P$ that matches current history. The probability of length 2 is the probability of choosing a history not the same as the current one in the second time step and getting the current history again in the third time step. It is $\frac{P-1}{P} \cdot \frac{1}{P}$. Similarly, we can obtain the probability of length $x$:

$$p_{rd}(x) = (\frac{P-1}{P})^{x-1}(\frac{1}{P}) \tag{3}$$

At each of $2P$ time steps we can check whether a length-1 loop will appear. However, only the first $2P - x$ time steps are possible to be the beginning of a length-$x$ loop. Thus the length distribution is given by

$$g_{rd}(x) = (2P - x) \cdot p_{rd}(x) = 2(1 - \frac{x}{2P})(1 - \frac{1}{P})^{x-1} \tag{4}$$

The average length for random rule can be found for large $P$ to be

$$\bar{x}_{rd} = \int_0^{2P} x g_{rd}(x)\, dx \approx \frac{4}{e^2 + 1} P \approx 0.477 P \tag{5}$$

When the standard updating rule is used, each history is followed by one of two fixed histories (Fig. 1a). If we order all histories according to their decimal values $i = 0, 1, 2, \cdots, P-1$, we see that the $i$-history is followed by either $2i \bmod P$ or $2i + 1 \bmod P$(Fig. 1a). If we start from $i$-history, what is the probability that the $i$-history will appear again after $x$ time steps[5]? A moment's reflection shows that the final history is given by $2^x i + j \bmod P$, with $j$ ranging from 0 to $2^x - 1$. Assume there are $s$ final

values among all $2^x$ possible values satisfying $2^x i + j \mod P = i$, which is equivalent to

$$(2^x - 1)i + j \mod P = 0 \tag{6}$$

For the small $\alpha$ case, because its quasi-periodic behavior, most histories appear the same times within first 2$P$ time steps (Fig. 4). It is thus a good approximation that the probability of length $x$ is related to the average of $s$ over all possible $i$, denoted by $\bar{s}(x)$. Now, $i$ can be one of 0, 1, 2, …, $P$-1, so the value $(2^x - 1)i + j$ can be any integer in 0, 1, 2, …, $(2^x - 1)P$. It is easy to see that there are $2^x$ values among them are multiple of $P$. So we have $\bar{s}(x) = 2^x / P$. Notice these $2^x$ solutions for length $x$ have included all solutions of smaller lengths which are factors of $x$. So we have to subtract them from $2^x$ in obtaining the probability of length $x$. For example, the probabilities of length 1 to length 6 are given below.

$$\begin{aligned}
p_{st}(1) &= \frac{1}{2^1 P}(2^1) = \frac{1}{P} \\
p_{st}(2) &= \frac{1}{2^2 P}(2^2 - 2^1) = \frac{1}{2P} \\
p_{st}(3) &= \frac{1}{2^3 P}(2^3 - 2^1) = \frac{3}{4P} \\
p_{st}(4) &= \frac{1}{2^4 P}(2^4 - 2^2) = \frac{3}{4P} \\
p_{st}(5) &= \frac{1}{2^5 P}(2^5 - 2^1) = \frac{15}{16P} \\
p_{st}(6) &= \frac{1}{2^6 P}(2^6 - 2^3 - 2^2 + 2^1) = \frac{27}{32P}
\end{aligned} \tag{7}$$

These values are consistent with the numerical results shown in inset of Fig. 3b. The probability of length $x$ is close to 1/$P$ for large $x$ because the contributions from its factors are very small comparing to $2^x$. To obtain the length distribution one has to take two factors into consideration. First, the probability $p_{st}(x)$ has to be weighted by the factor $2P - x$ as explained in the random case. Second, in the standard updating rule, when a history appears first time, it is followed by one of two possible histories. In the case of small $\alpha$ that we consider here, when the history appears the second time, it will be followed by the other history of the two[16,12]. That is, two appearances of a history compensate for each other in determining the probability of finding a loop of length $x$. Thus we have to weight the distribution by a factor 1/2. The length distribution is then given by

$$g_{st}(x) = \frac{1}{2} \cdot (2P - x) \cdot p_{st}(x) \underset{x \gg 1}{\approx} 1 - \frac{x}{2P} \tag{8}$$

The average length for the standard rule is

$$\bar{x}_{st} = \int_0^{2P} x g_{st}(x)\, dx \approx \frac{2P}{3} \tag{9}$$

From Eqs. (5) and (9), we have $\bar{x}_{rd} < \bar{x}_{st} < \bar{x}_{pd} = P$, with its order consistent with order of $\chi_{rd}, \chi_{st}, \chi_{pd}$ shown in Fig. 2.

To calculate $\chi_{rd}$ and $\chi_{st}$ according to Eq. (1), we need to know the distribution of $l$ for standard and random rules: $h_{st}(l)$ and $h_{rd}(l)$. Results of simulations, averaging 1000 runs, are plotted in Fig. 5. The values for $\chi_{rd}$ and $\chi_{st}$ are therefore given by

$$\chi_{st} - \frac{1}{8} = \frac{c_S}{\alpha} \sum_{l=1}^{P} \frac{h_{st}(l)}{l} \equiv \frac{c_S}{\alpha} (\log P'_{st} + \gamma) \tag{10}$$

and

$$\chi_{rd} - \frac{1}{8} = \frac{c_S}{\alpha} \sum_{l=1}^{P} \frac{h_{rd}(l)}{l} \equiv \frac{c_S}{\alpha} (\log P'_{rd} + \gamma) \tag{11}$$

From Eqs. (10), (11), we can determine effective information for standard and random rules. Results of $P'_{rd}$ and $P'_{st}$ for some different values of $P$ are listed in Table 1. For fixed $N$, $P$ can not be too small in order the statistical analysis to be valid. $P$ can not be too large either so that $\alpha$ will not leave symmetric phase[17].

Table 1   Effective information

($N = 20001$, $S = 2$, average of 5000 runs)

| $P$ | 16 | 32 | 64 | 128 | 256 | 512 | 1024 |
|---|---|---|---|---|---|---|---|
| $P'_{pd}$ | 16 | 32 | 64 | 128 | 256 | 512 | 1024 |
| $P'_{rd}$ | 3.9 | 4.0 | 4.1 | 4.1 | 4.1 | 4.1 | 4.1 |
| $P'_{st}$ | 5.0 | 6.4 | 8.0 | 10.4 | 14.0 | 18.9 | 24.2 |

As a test for our values of effective memories given in Table 1, in Fig. 6 we plot $\chi_{rd}$ and $\chi_{st}$ versus $\alpha$ for $P = 64$. The former coincides very well with the curve of $\chi_{pd}$ for $P = 4$ as predicted in Table 1. The latter coincides better with the curve of $\chi_{pd}$ for $P = 10$ instead of $P = 8$ as predicted in Table 1. Qualitatively, the calculations of the effective memory are satisfactory.

From Table 1 we found the results of effective information are very interesting. The effective information for random rule $P'_{rd}$ is almost independent of $P$. In terms of effective memory, $M'_{rd} \approx 2$ for any $M$. On the other hand, the effective information for standard rule $P'_{st}$ is a monotonic function of $P$. The corresponding effective memory has a linear relation with $M$: $M'_{st} = 0.384M + 0.742$ (Fig. 7). Further investigation is in order.

## V.   Conclusion

We have simulated the minority game in the symmetric phase. The results showed that the variance of winning population $\chi$ in this phase is approximately inverse-proportional to information per agent per strategy $\alpha$. We used three kinds of rules-- standard, random, and periodic rules--to update the history and found their results were qualitatively similar but differed in quantity substantially in the symmetric phase. We defined the effective memory for each updating rule based on a previous analysis to quantify their differences and showed how the results were dependent on the memory.

**Figure captions**

Fig. 1

(a) Standard game uses de Bruijn graph to update its history. A history $i$ can be updated to history $2i$ or $2i+1 \mod P = 2^M$. Here $M = 3$.

(b) In the periodic rule, a history $i$ is updated to $i+1 \mod P$ regardless of the winning results.

Fig. 2

Variance per agent $\chi$ versus information per agent $\alpha$ using three different updating rules: periodic(triangle), standard(circle), and random(square). These curves show distinct behavior in each of the three regions separated at $\alpha_{c_1}$ and $\alpha_{c_2}$.

Fig. 3

Histogram(gray) of the length-distribution of the loops in a quasi-period $2P$ for (a) random rule and (b) standard rule. Thick lines are analytical results. Inset of Fig. 3(b) shows details of numerical results (black) for 6 short lengths. Analytical results (white) of Eq. (7) are also shown for comparison.

Fig. 4

A typical plot of appearing frequency versus history in a quasi-period $2P$ for standard rule.

Fig. 5

Distributions of the number of histories($l$) that appear odd times when a loop is found for random rule(square) and standard rule(circle).

Fig. 6

$\chi$ versus $\alpha$ in the symmetric region. Three solid lines are results using periodic rule with $P = 64$, 10, 4. Simulation results of standard rule(circle) and random rule(triangle) with $P = 64$ show to have effective information $P'_{st} = 10$ and $P'_{rd} = 4$ respectively.

Fig. 7

Effective memory of standard rule $M'_{st}$ has a linear relation with the memory $M$.

Fig. 1a

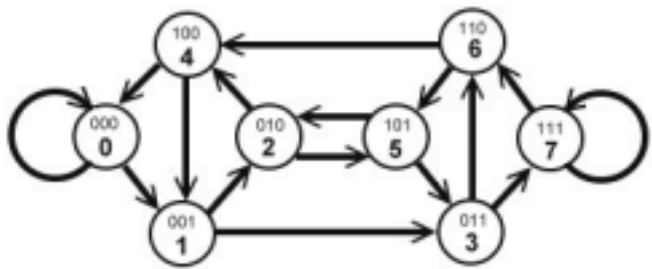

Fig. 1b

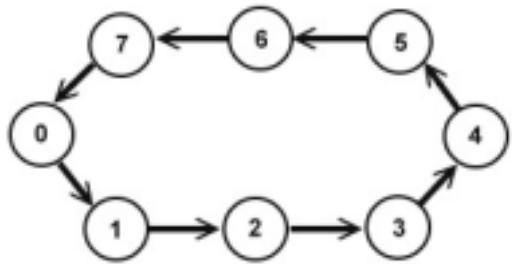

Fig. 2

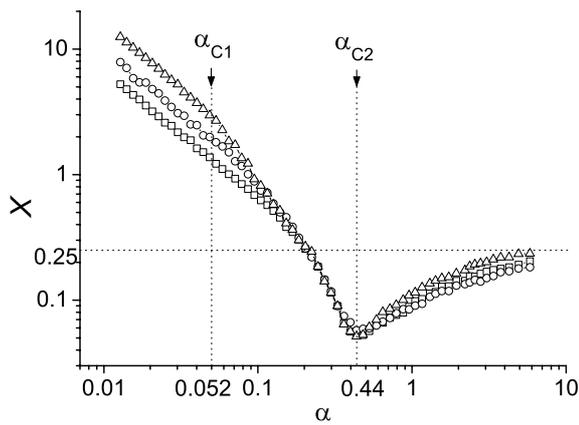

Fig. 3a                    Fig. 3b

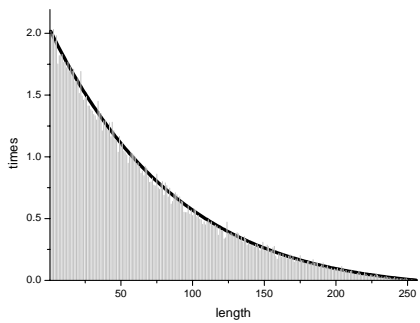 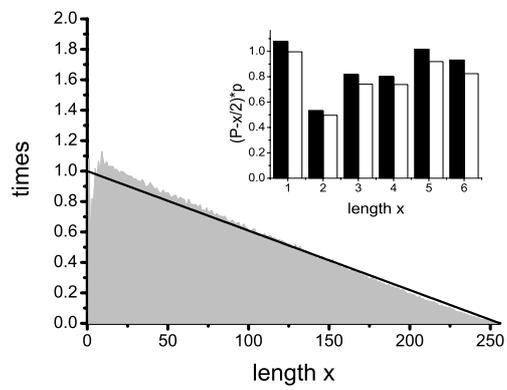

Fig.4

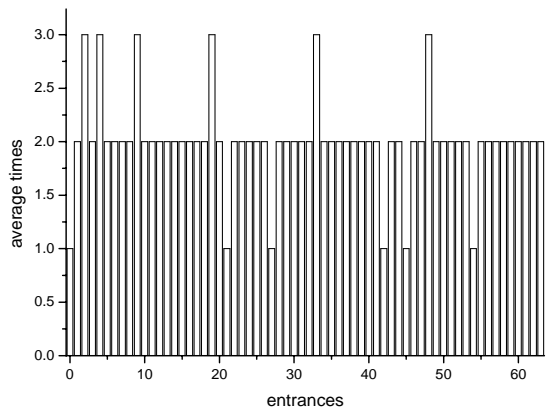

Fig. 5

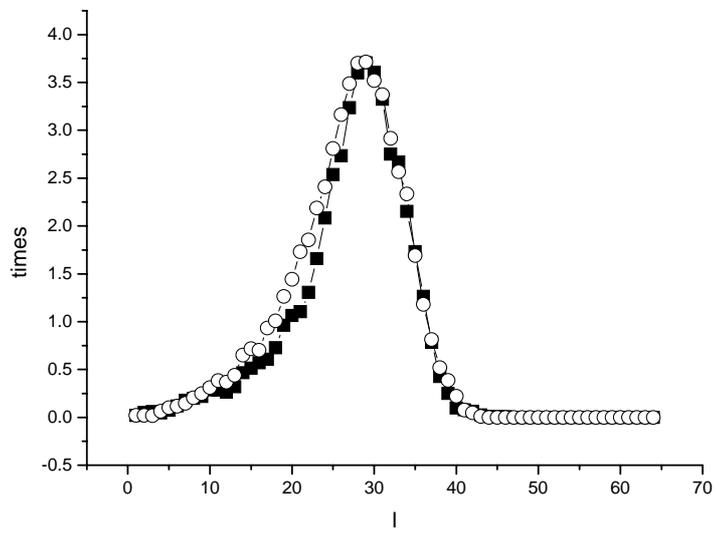

Fig. 6

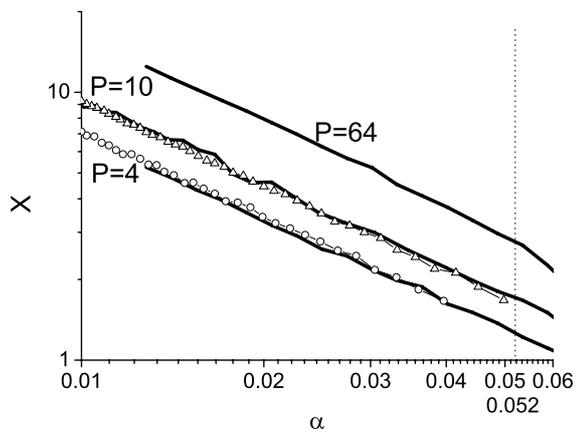

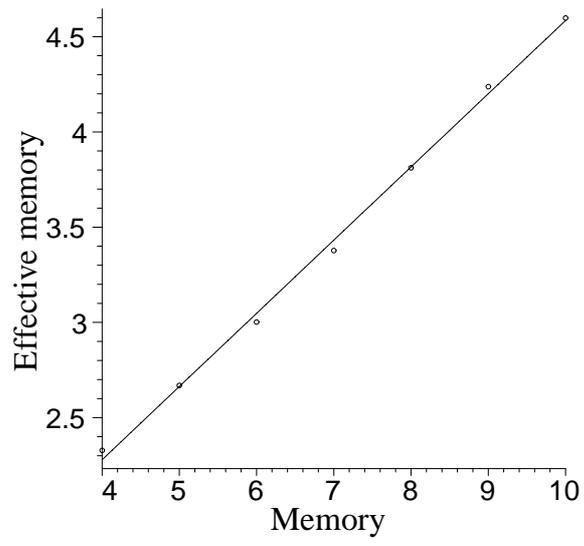

Fig. 7